\documentclass[fleqn,usenatbib,useAMS]{mnras}
\usepackage{amsmath}
\usepackage{mathrsfs}
\usepackage{amsfonts}
\usepackage{graphicx}
\usepackage{color}
\usepackage{mathtools}
\usepackage{subfigure}
\usepackage[T1]{fontenc}

\title[Searching for Cross-Correlation Between SGWB and Galaxy Number Counts]{Searching for Cross-Correlation Between Stochastic Gravitational Wave Background and Galaxy Number Counts}

\author[K.Z.Yang et al.]{Kate Z.Yang, $^{1}$\thanks{E-mail:yang5991@umn.edu} Vuk Mandic$^{1}$, Claudia Scarlata$^{1}$ and Sharan Banagiri$^{1}$
\\
$^{1}$School of Physics and Astronomy, University of Minnesota, Minneapolis, MN 55455, USA}

\pubyear{2020}
\begin{document}
\maketitle

\begin{abstract}
Advanced LIGO and Advanced Virgo have recently published the upper limit measurement of persistent directional stochastic gravitational wave background (SGWB) based on data from their first and second observing runs (O1 and O2) \citep{LIGOScientific:2019gaw}. In this paper we investigate whether a correlation exists between this maximal likelihood SGWB map and the electromagnetic tracers of matter structure in the universe, such as galaxy number counts. The method we develop will improve the sensitivity of future searches for anisotropy in the SGWB and expand the use of SGWB anisotropy to probe the formation of structure in the universe. In order to compute the cross-correlation, we used the spherical harmonic decomposition of SGWB in multiple frequency bands and converted them into pixel-based sky maps in HEALPix \citep{Gorski:1999rt} basis. For the electromagnetic (EM) part, we use the Sloan Digital Sky Survey (SDSS) galaxy catalog and form HEALPix sky maps of galaxy number counts at the same angular resolution as the SGWB maps. We compute the pixel-based coherence between these SGWB and galaxy count maps. After evaluating our results in different SGWB frequency bands and in different galaxy redshift bins, we conclude that the coherence between the SGWB and galaxy number count maps is dominated by the null measurement noise in the SGWB maps, and therefore not statistically significant. We expect the results of this analysis to be significantly improved by using the more sensitive upcoming SGWB measurements based on the third observing run (O3) of Advanced LIGO and Advanced Virgo.
\end{abstract}
\section{Introduction}\label{sec:intro}
Recent detections of gravitational waves (GWs) from the mergers of binary black hole (BBH)~\citep{Abbott:2016drs,Abbott:2016nmj,Abbott:2017vtc,Abbott:2017gyy,Abbott:2017oio,LIGOScientific:2018mvr,Salemi:2019ovz} and binary neutron star (BNS)~\citep{Abbott:2017ntl} systems have initiated the field of multi-messenger astrophysics. These discoveries have triggered a broad range of studies including novel tests of General Relativity \citep{TheLIGOScientific:2016src,Monitor:2017mdv}, constraints on the neutron star equation of state~\citep{Abbott:2017ntl}, estimates of the BBH and BNS rates and studies of their progenitors~\citep{Abbott:2016drs,TheLIGOScientific:2016htt,Abbott:2017ntl}, a new measurement of the Hubble constant $H_0$~\citep{Abbott:2017xzu}, kilonova interpretation of the observed BNS merger~\citep{Abbott:2017wuw}, and others.
The observed rates of resolved BBH and BNS mergers also imply a relatively strong stochastic gravitational-wave background (SGWB) arising from the signal superposition of such mergers throughout the universe~\citep{TheLIGOScientific:2016wyq,Abbott:2017xzg,LIGOScientific:2019vic}.

As the ground-based Advanced LIGO (aLIGO)~\citep{TheLIGOScientific:2014jea} and Advanced Virgo (aVirgo)~\citep{TheVirgo:2014hva} GW detectors improve their strain sensitivities, one of their primary targets will be the detection of this SGWB~\citep{TheLIGOScientific:2016xzw,LIGOScientific:2019vic,LIGOScientific:2019vic}. The projected strain sensitivity improvements~\citep{Aasi:2013wya} combined with the recently proposed Bayesian search technique~\citep{Smith:2017vfk} have made the SGWB detection with advanced detectors a real possibility. It is worth noting that SGWB can also be generated through a variety of stochastic processes in the early Universe~\citep{Romano:2016dpx}, including models of inflation, phase transitions, cosmic strings, and others. 

Due to this discovery potential, there has been a recent surge in the literature studying the possible anisotropy of the SGWB due to BBH and BNS mergers~\citep{Contaldi:2016koz,Jenkins:2018uac,Jenkins:2018kxc,Jenkins:2019uzp,Jenkins:2019nks,Bertacca:2019fnt,Cusin:2017fwz,Cusin:2017mjm,Cusin:2018rsq,Cusin:2018ump, Cusin:2019jpv,Pitrou:2019rjz,Canas-Herrera:2019npr,Stiskalek:2020wbj,Cavaglia:2020fnc, Payne2006}, as well as in cosmological SGWB models due to phase transitions~\citep{Geller:2018mwu} and cosmic strings~\citep{Jenkins:2018lvb}. Several contributions have also investigated the possibility of correlating the SGWB anisotropy with the anisotropy observed in EM tracers of the large scale structure, such as galaxy counts and weak lensing~\citep{Cusin:2017fwz,Cusin:2017mjm,Cusin:2018rsq,Cusin:2019jpv,Canas-Herrera:2019npr,Scelfo:2018sny,Oguri:2016dgk,Oguri:2016dgk,Mukherjee:2019wcg,Mukherjee:2018ebj}, or the Cosmic Microwave Background~\citep[CMB, ][]{Geller:2018mwu}. The first theoretical predictions of the cross-correlation power spectrum between the SGWB and the galaxy number counts have been made, as well as between the SGWB and the weak lensing convergence, including the dependence on SGWB frequency and the galaxy redshift distribution~\citep{Cusin:2018rsq}.

Correlating the SGWB anisotropy with anisotropy in EM tracers offers multiple lines of inquiry. First, the GW-EM correlation method is likely to be more sensitive when trying to detect the SGWB anisotropy than the traditional techniques that rely on GW data alone. Second, the GW-EM correlation can be placed in a parameter estimation framework so as to measure the cosmological and astrophysical parameters of the model that gives rise to the SGWB anisotropy~\citep{Cusin:2017fwz,Cusin:2018rsq,Cusin:2019jpv,Canas-Herrera:2019npr,Mukherjee:2019oma}--for example, to constrain the formation and evolution of structure in the universe. Third, correlations with different EM tracers (e.g. galaxy counts vs CMB) may enable separating different SGWB contributions (e.g. due to binary mergers vs cosmological, respectively). 

In this paper, we present the first analysis of the GW-EM anisotropy correlations, using the data from the second observation run of Advanced LIGO and Advanced Virgo correlated with the galaxy count survey from the Sloan Digital Sky Survey (SDSS). We observe no significant correlations in the data and hence place upper limits on the correlation parameter. The rest of this paper is organized as follows. In Section 2 we review the method for measuring the SGWB anisotropy and we apply it to different frequency bands of Advanced LIGO data to compute SGWB sky maps at different angular resolutions. In Section 3 we review the SDSS survey and compute the maps of the galaxy count distribution across the sky in several redshift slices. In Section 4 we compute the correlations between the SGWB and SDSS maps and establish the first upper limits on the correlation coefficients. In Section 5 we offer our concluding remarks and discuss the numerous ways of extending our study in the future. 
\section{SGWB Anisotropy Upper Limit}\label{sec:ligo}
Stochastic gravitational-wave background arises as a superposition of waves from many incoherent GW sources. The SGWB therefore does not have a deterministic waveform, and is instead characterized by its energy density spectrum. In particular, we define the frequency and angular GW energy density spectrum $\Omega_{\rm{GW}}(f,\Theta)$ as:
\begin{align}
		\Omega_{\rm{GW}}(f,\Theta)=\frac{f}{\rho_c}\frac{d^3 \rho_{\rm{GW}}}{df d^2 \Theta},
\end{align}
where $\rho_{\rm{GW}}$ is the GW energy density, $f$ is frequency, $\Theta$ represents a direction on the sky, and $\rho_c$ is the critical energy density needed to close the universe. Past searches for the SGWB anisotropy have assumed that this spectrum can be factorized into frequency and sky-direction parts~\citep{Thrane:2009fp, LIGOScientific:2019gaw, TheLIGOScientific:2016xzw}:
	\begin{align}
		\Omega_{\rm{GW}}(f,\Theta)=\frac{2\pi^2}{3H_0^2}f^3 H(f){\mathcal{P}}(\Theta),
		\label{Eq:omega}
	\end{align}
where $H_0$ is the Hubble constant, ${\mathcal{P}}(\Theta)$ captures the angular dependence on the sky, and $H(f)$ describes the frequency dependence of the spectrum, typically assumed to take a power law form: $H(f) = (f/f_{\rm ref})^{\alpha - 3}$, with some reference frequency $f_{\rm{ref}}$ and spectral index $\alpha$. For the SGWB due to BBH and BNS mergers $\alpha = 2/3$~\citep{TheLIGOScientific:2016wyq,Abbott:2017xzg}, but different values of the spectral index are appropriate for other models. In this paper we adopt $f_{\rm ref} = 100$~Hz.

The spatial dependence can be decomposed into any set of basis functions on a sphere--we will use the spherical harmonics basis:
	\begin{align}
		P(\Theta)=\sum_{l m} P_{l m}Y_{l m}(\Theta)
	\end{align}
The objective of the SGWB anisotropy upper limit analysis is therefore to estimate the parameters $P_{lm}$. We adopt the approach developed in \citep{Thrane:2009fp} and used in past anisotropic SGWB searches~\citep{LIGOScientific:2019gaw, TheLIGOScientific:2016xzw}, which starts with the cross-correlation between the strain time series data of GW detectors (LIGO Hanford (H1) and LIGO Livingston (L1) in our case): 
	\begin{align}
		C(f,t) = s_1^*(f,t)\ s_2(f,t), 
	\end{align}
where $t$ denotes a time segment, and $s_1$ and $s_2$ are Fourier transforms of the strain time series of H1 and L1 in this time segment. We then define \textit{the dirty map} $X_\nu$:
	\begin{align}
		X_\nu=\sum_{f,t}\gamma_\nu^* (f,t)\frac{H(f)}{P_1(f,t) P_2(f,t)}C(f,t),
	\end{align}
where the sum is over all frequency bins $f$ and all time segments $t$. The index $\nu$ runs over the spherical harmonic components (i.e. $\nu \equiv (l,m)$), $P_{1,2}(f,t)$ are strain power spectral densities for the two detectors, and $\gamma_{\nu}(f,t)$ is a geometric factor that is a function of the separation and relative orientation of the LIGO detectors H1 and L1~\citep{Christensen:1992wi,Thrane:2009fp}.

The dirty map $X_\nu$ represents an estimate of the GW energy density sky distribution convolved with the response of the detectors' antenna patterns. The corresponding uncertainty is described by the covariance matrix, also known as the \textit{Fisher matrix}:
	\begin{align}
		\Gamma_{\mu \nu} = \sum_{f,t} \gamma_\mu^*(f,t) \frac{H^2(f)}{P_1(f,t) P_2(f,t)}\gamma_\nu(f,t).
	\end{align}
Estimators of the spherical harmonic coefficients $P_{lm}$, also known as the \textit{the clean map}, are then given by~\citep{Thrane:2009fp}:
	\begin{align}\label{eq:p0}
		\hat{P}_\mu=\sum_{\nu} \big(\Gamma_R^{-1}\big)_{\mu\nu}X_\nu.
	\end{align}
The covariance matrix corresponding to the $P_{lm}$'s is the inverse of the Fisher matrix, $\Gamma_R^{-1}$. In general, the Fisher matrix may be singular, reflecting the fact that the GW detector network may be insensitive to some directions on the sky. The inversion of this matrix therefore requires regularization, which is accomplished by diagonalizing the Fisher matrix and removing the eigenvalues that are close to zero (i.e. typically setting about 1/3 of the lowest eigenvalues to infinity)~\citep{Thrane:2009fp}. The subscript $R$ in $\Gamma_R^{-1}$ denotes that the Fisher matrix has been regularized. The regularization does not induce a bias on the clean map $\hat{P}_\mu$.

The angular resolution of this technique is set by a diffraction-like limit~\citep{Thrane:2009fp}:
	\begin{align}
	\theta=\frac{c}{2df}\approx \frac{50 \rm{Hz}}{f_\alpha},
	\end{align}
where $\theta$ is in radians, $d$ is the distance between H1 and L1 (3000 km), and $f_\alpha$ is typically taken to be the most sensitive frequency in the detector band for a power law SGWB with spectral index $\alpha$, and for the given detector noise power spectra \citep{TheLIGOScientific:2016xzw, Romano:2016dpx}. For the BBH/BNS background, $\alpha=2/3$ and the most sensitive frequency in the past searches was found to be $50-60$ Hz, implying a coarse angular resolution of order $\theta \approx \pi/3$ and therefore spherical harmonic decomposition up to $l_{\rm max} = \pi / \theta \approx 3-4$~\citep{TheLIGOScientific:2016xzw,LIGOScientific:2019gaw}. 

In an attempt to probe finer angular scales, we will conduct the above analysis in several narrower frequency bands: 50-100 Hz, 100-150 Hz, 150-200 Hz, 200-250 Hz. The higher frequency bands will result in better angular resolution, specifically in $l_{max} = $ 4, 8, 12, and 16, respectively. 
For the highest $l_{max}$=16 the corresponding angular resolution is $\theta \approx$ 7.3 deg. We note, however, that the sensitivity of the search is reduced at higher frequencies, both because of the poorer strain sensitivity of the GW detectors above $\sim 100$ Hz~~\citep{TheLIGOScientific:2014jea,TheVirgo:2014hva} and because of the $f^3$ term in Eq.~\ref{Eq:omega}.

\begin{figure*}
\centering
\begin{subfigure}
    \centering
    \scalebox{0.26}{\includegraphics{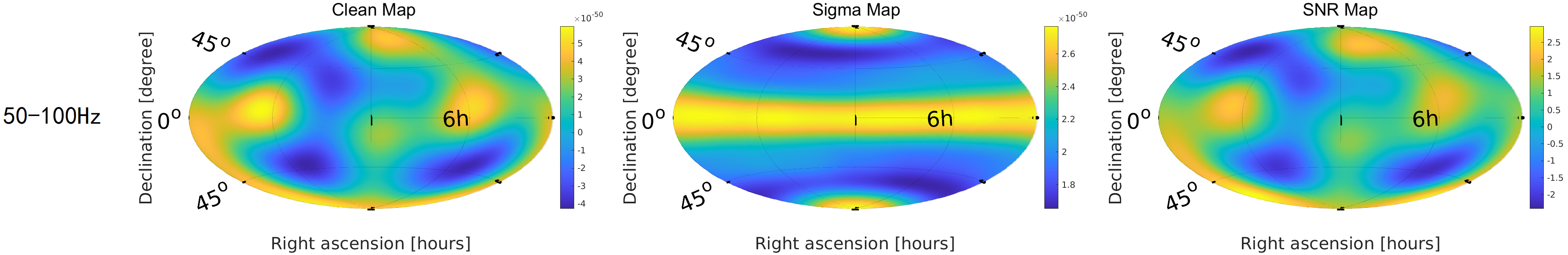}}
\end{subfigure}
\begin{subfigure}
    \centering
    \scalebox{0.26}{\includegraphics{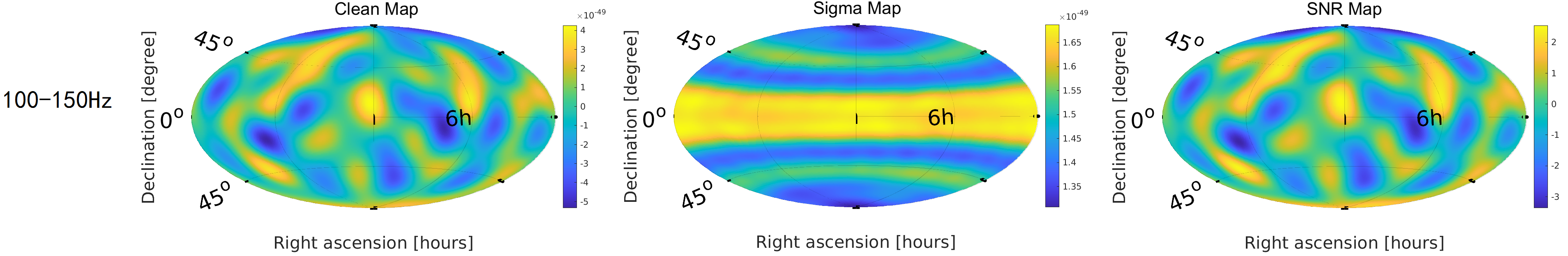}}
\end{subfigure}
\begin{subfigure}
    \centering
    \scalebox{0.26}{\includegraphics{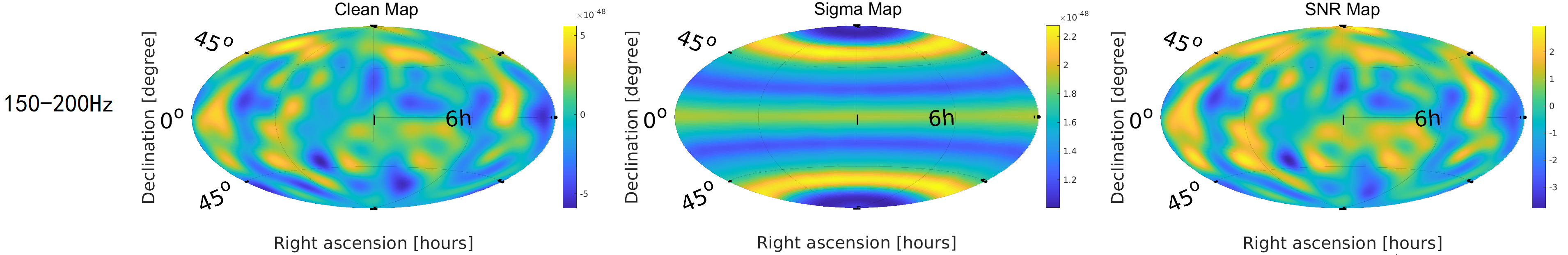}}
\end{subfigure}
\begin{subfigure}
    \centering
    \scalebox{0.26}{\includegraphics{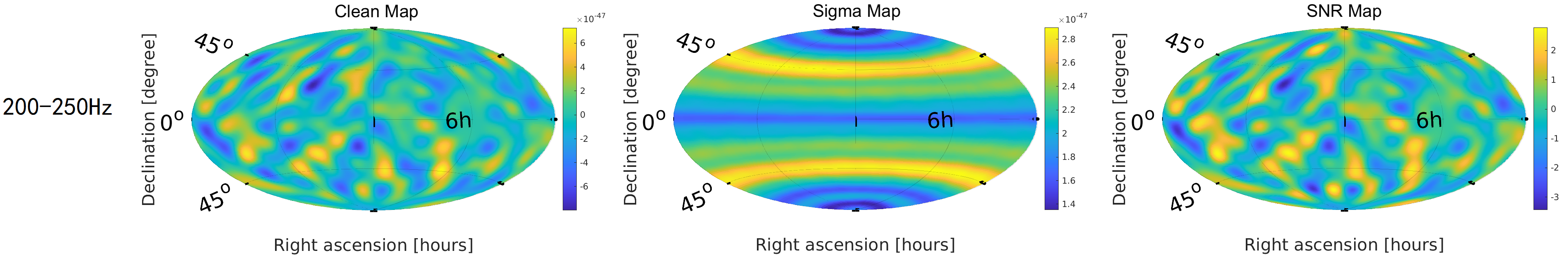}}
\end{subfigure}
\caption{LIGO O2 clean maps, sigma maps and signal-to-noise (SNR) maps for the four frequency bands of 50-100, 100-150, 150-200, 200-250 Hz from top to bottom, with $\alpha$ of 2/3 and $l_{max}$ of 4, 8, 12 and 16, respectively.}
\label{fig:O2-stochastic}
\end{figure*}

We apply the above analysis procedure to the GW data from the second observing run (O2) of Advanced LIGO’s detectors H1 and L1. The O2 data are collected from 16:00:00 UTC on 30 November, 2016 to 22:00:00 UTC on 25 August, 2017 \citep{LIGOScientific:2019gaw}. We follow closely the data processing procedure described in \citep{TheLIGOScientific:2016xzw,LIGOScientific:2019gaw}. The time-series data are divided into 50\% overlapping segments of 192 seconds, passing through a cascading high-pass filter. The data segments are then Fourier transformed into the frequency domain, and the H1-L1 cross-correlation is computed for each 192 second long segment. The results from these overlapping time segments are optimally combined to produce the final cross-correlation estimate. We use the same data selection criteria described in \citep{LIGOScientific:2019gaw}. Finally we compute the clean map estimates following Eq.~\ref{eq:p0}, for each of the four frequency bands. The resulting clean maps, sigma maps and signal-to-noise (SNR) maps for the four frequency bands are shown in Figure \ref{fig:O2-stochastic}.
\section{Galaxy Count Anisotropy}\label{sec:sdss}
As an example of an EM tracer of matter structure, we will use the distribution of galaxy counts across the sky. The most complete and largest area galaxy survey currently comes from the Sloan Digital Sky Survey (SDSS), whose Data Release~16 (DR16) contains observations through August 2018 \citep{SDSS_DR16}. The SDSS imaging data contain observations covering almost 1.5$\times 10^4$~deg$^2$ or roughly 1/3 of the sky. The photometric catalog includes approximately 2$\times 10^8$ galaxies with r-band magnitude brighter than $m_r \approx 22.2$. 
In addition to the imaging observations, SDSS acquired spectra for $\approx 1.8\times 10^6$ galaxies brighter than $m_r\approx 17.7$. For galaxies fainter than this limit, SDSS provides an estimate of the galaxy photometric redshift based on the analysis of the five photometric bands (hereafter, photo-$z$). Although the resulting redshifts are substantially more uncertain than those derived from spectroscopic observations, the use of photo-$z$ allows us to increase the sample size considerably.

From the SDSS archive, we select all {\it galaxies} with magnitudes in the 17$< m_r \le 21$ range. To identify only galaxies, we use the SDSS \emph{type} parameter ($type = 3$). The magnitude range was chosen to ensure a survey completeness level of 90\% or better, and to minimize the contamination to the galaxy sample by misclassified stars \citep[see ][for a discussion]{Wang:2013noa}. We also constrain the analysis to include data in a fully contiguous area mostly in the northern Galactic hemisphere.
The final photometric galaxy catalog includes 23 million objects with median photometric redshift of 0.33. 
For a subsample of 1.4 million galaxies, spectroscopic redshifts are available, with a median spectroscopic redshift of 0.39.

A number of systematic effects can potentially affect the spatial distribution of galaxies on the large scales relevant for the cross-correlation with the GW maps. Here we consider only the effects of atmospheric seeing variation and Milky Way extinction, as they impact the observed galaxy number counts on degree scales and above \citep{Reid:2015gra, Ross:2016gvb}. We follow \citep{Wang:2013noa} and we consider in the analysis only areas of good seeing and minimal Galactic extinction. We quantify the seeing using the average Full Width Half Maximum (FWHM) of the point spread function (PSF) during the observations and exclude from the analysis those sky regions with average FWHM $\ge 1.5 ''$. This cut is found to exclude $\sim$ 12\% of the total area. Galactic extinction is characterized via the color excess, E(B-V), and we exclude areas with $E(B-V)> 0.13$, or 15\% of the total area.
The seeing and galactic extinction cuts can have significant effects on the average number of galaxies in some areas of the sky. In order to obtain unbiased galaxy count maps, we apply the following procedure. In the HEALPix basis, the full sky is divided into pixels of the same angular size ~\citep{Gorski:1999rt}, a convenient choice for the computations of cross correlations with the GW sky maps. The number of pixels in the HEALPix basis is chosen to match the value of $l_{\rm max}$ for each frequency band:
	\begin{align}
		{\rm{\# pixels }}\approx \frac{4\pi}{\theta^2}=\frac{4}{\pi}\cdot l_{max}^2.
	\end{align}

While we ultimately need a galaxy count map of resolution corresponding to $l_{\rm max}=16$, corresponding to an angular scale of $\sim 7^{\circ}$, we start by producing the HEALPix map for the SDSS photometric catalog with a higher resolution (small pixels). The small pixels have an angular scale of 2.4$^{\circ}$. This angular scale is small enough that the seeing and galactic extinction do not vary too much for the galaxies within the pixels, but large enough to ensure a large number of galaxies (on average $>10^2$ galaxies). Using all galaxies in each small pixel, we compute the average r-band seeing and extinction for that pixel. We reject all galaxies within a pixel whose average seeing is greater than 1.5 arcsecond or the galactic extinction is $>$ 0.13.

\begin{figure}
	\subfigure[]
	{\scalebox{0.3}{\includegraphics{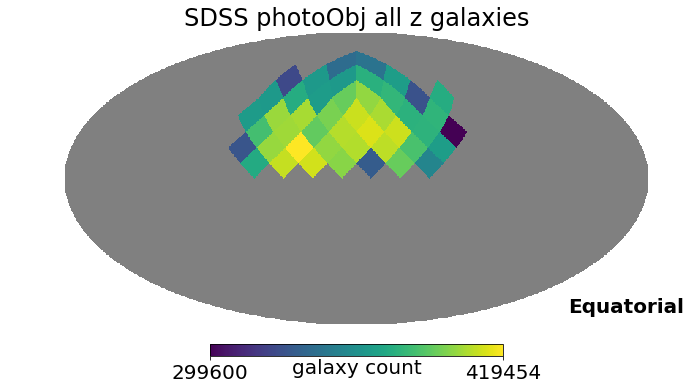}}}
	\subfigure[]
	{\scalebox{0.3}{\includegraphics{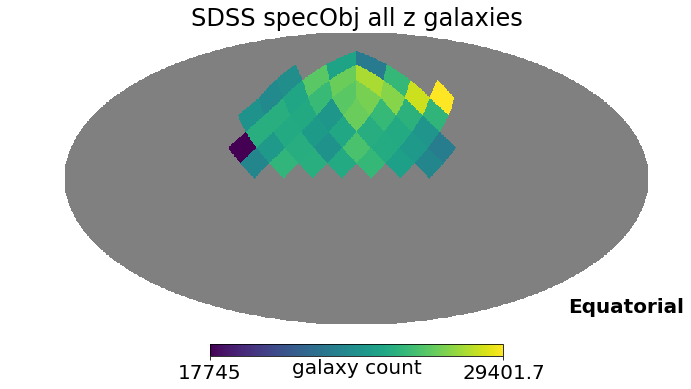}}}
	\caption{HEALPix sky maps of SDSS galaxy number counts for an angular resolution corresponding to $l_{max}=16$, after applying the data quality cuts described in the text. (a) and (b) depict the photometric and spectroscopic catalogs respectively. The color scales with the galaxy count in each pixel.}
	\label{fig:sdssweighted}
	\end{figure}

To correct for the missing pixels we then replace the counts in that rejected pixel with the average counts of the other small pixels inside a larger HEALPix pixel corresponding to $l_{\rm max} = 16$. Since all objects in the SDSS spectroscopic catalog are also in the SDSS photometric catalog, we apply the same procedure to the spectroscopic catalog as well. The results for both catalogs are shown in Figure \ref{fig:sdssweighted}.
%

Since the SGWB due to BBH and BNS mergers at different angular scales is expected to be dominated by binaries at different redshifts~\citep{Cusin:2019jpv}, we will conduct our analysis in several redshift bins, i.e. compute the correlation between SGWB sky maps and the galaxy number sky maps in each redshift bin, respectively. We choose to divide both catalogs into redshift bins of width 0.1 (i.e. 0.0-0.1, 0.1-0.2,...). For the photometric catalog, we extend the analysis up to redshift 0.6, which includes 97\% of all the galaxies. For the spectroscopic catalog, we go up to 0.7 and the redshift slicing includes 98\% galaxies.
While the photometric and spectroscopic catalog maps including all redshifts do not appear to be correlated (as shown in Figure \ref{fig:sdssweighted}), we have confirmed that the photometric and spectroscopic maps in each redshift bin are highly correlated.
\section{SGWB-EM Correlations}\label{sec:corr}
Having produced the sky maps for the SGWB in each of the four frequency bands and for the galaxy counts in each of the redshift bins, we next compute the correlations between these maps. Denoting the SGWB energy density in a pixel $i$ as $M_{{\rm GW},i}$ and the galaxy number count in the same pixel as $M_{{\rm GC},i}$, we define the corresponding fluctuations:
	\begin{align}
	\delta M_{{\rm GW},i} = M_{{\rm GW},i} -\langle M_{\rm GW} \rangle, \ \ \delta M_{{\rm GC},i} = M_{{\rm GC},i} -\langle M_{\rm GC} \rangle.
	\end{align}
Then we define the coherence between these fluctuations as:
	\begin{align}
	\Gamma=\frac{\langle \delta M_{\rm GW} \cdot \delta M_{\rm GC} \rangle ^2}{\langle \delta M_{\rm GW}^2\rangle \langle \delta M_{\rm GC}^2\rangle}.
	\end{align}
The averages are computed over all pixels in the maps. 

To assess the significance of the measured coherence, we use simulations. In particular, we generate 10,000 simulated SGWB noise maps assuming zero-mean multivariate Gaussian distribution for $P_{lm}$'s described by the regularized inverse Fisher matrix obtained from LIGO data (see Section \ref{sec:ligo}). We then compute $\Gamma$ for these simulated maps and the galaxy count sky maps in different redshift bins respectively. We then compute the false alarm rate (FAR) as a function of coherence:
	\begin{align}
	{\rm{FAR}} (\Gamma) =\frac{{\rm{\#\ of\ events}} >\Gamma}{{\rm{total\ \#\ of\ events}}}.
	\label{eq:far}
	\end{align}

\begin{figure}
		 \scalebox{0.6}{\includegraphics{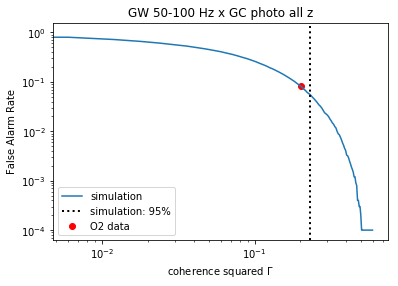}}
		\caption{The False Alarm Rate of the coherence squared $\Gamma$ distribution of 10,000 simulated SGWB sky maps in the 50-100 Hz band and the galaxy count sky maps from the full photometric SDSS catalog (blue curve). The coherence squared $\Gamma$ obtained from the LIGO O2 data and the same galaxy map is represented as a red dot. The black vertical dashed line represents the 95 percentile of the 10,000 simulations.
	}
		\label{fig:pfar}
	\end{figure}

Figure \ref{fig:pfar} shows an example of the false alarm rate calculation for the specific case of the 50-100 Hz band SGWB map and the full photometric SDSS galaxy catalog. The blue curve is derived from the 10,000 simulations and the red dot denotes the actual measured coherence using the O2 LIGO data. The FAR value of the red dot then gives the p-value significance of the measured coherence.

\begin{figure}
	\subfigure[]
	{
		\scalebox{0.6}{\includegraphics{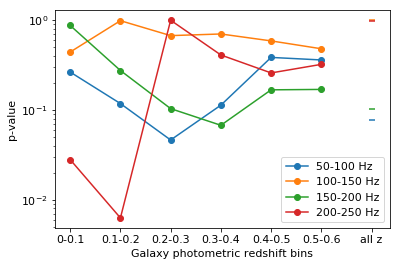}}
	}
	\subfigure[]
	{
		\scalebox{0.6}{\includegraphics{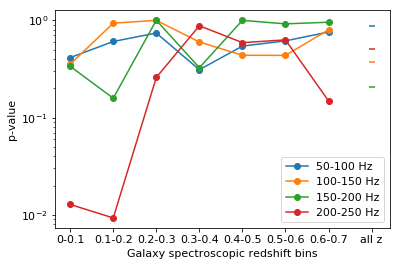}}
	}
		\caption{The p-value significance of the coherences between the LIGO O2 SGWB maps in different frequency bins and the SDSS galaxy count maps in different redshift slices are shown for the (a) photometric and (b) spectroscopic SDSS catalogs. Note that the red and orange underscore in (a) are partially overlapping. All p-values are larger than $10^{-1}$ - $10^{-2}$
		, indicating no evidence for correlations between the SGWB and galaxy count maps.
		}
		\label{fig:pvalue}
	\end{figure}

We repeat this procedure for all four frequency bands of the GW data and all redshift bins of the SDSS data, and for both the photometric and spectroscopic SDSS catalogs. The resulting p-values are shown in Figure \ref{fig:pvalue}. 
The p-values for the photometric catalog have wider spread than the spectroscopic catalog, which can be explained by the fact that the photometric catalog is more uncertain in redshift and therefore more noisy as galaxy count maps in redshift slices. 
Above all it is evident that all p-values are at or above $10^{-1}$ - $10^{-2}$ for both the photometric and spectroscopic galaxy count maps,
indicating low statistical significance. We therefore observe no correlation between SGWB and galaxy count sky maps.

We note that this analysis can be extended to perform model selection and/or parameter estimation. For this paper, we consider a simple empirical model where we assume that the SGWB energy density fluctuations are proportional to the normalized galaxy density fluctuations: 
\begin{align}
\delta M_{{\rm GW},i}^{\rm model} = \lambda\cdot \frac{\delta M_{{\rm GC},i}}{\langle M_{{\rm GC},i}\rangle} + \delta M_{\rm GW}^{noise} ,
\label{eq:lambda}
\end{align}
where the index $i$ =1,2 represents galaxy count maps of the photometric or spectroscopic catalogs in the full redshift range. The factor $\lambda$ can therefore be interpreted as the GW strain power 
per normalized fluctuation in the galaxy number count. We can use the observed value of $\Gamma$ to constrain the model parameter $\lambda$. To do so, we scan the values of the scaling parameter $\lambda$; for each value of $\lambda$ we generate 1000 realizations of the SGWB noise map $\delta M_{\rm GW}^{noise}$ similarly to above, and for each realization we compute the coherence $\Gamma$ between the corresponding model $\delta M_{{\rm GW},i}^{\rm model}$ and the galaxy count map $\delta M_{{\rm GC},i}$. 

\begin{figure}
\centering
\scalebox{0.6}{\includegraphics{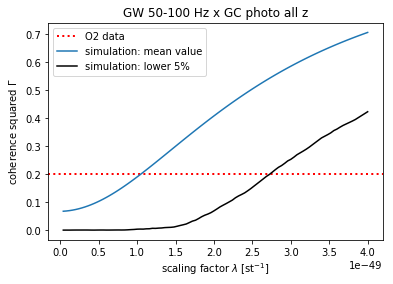}}
\caption{Coherence squared $\Gamma$ for simulated SGWB model maps described in the text is shown as a function of the scaling parameter $\lambda$, for the case of SGWB map in the 50-100 Hz band and for the full photometric SDSS catalog. The SGWB model maps were generated by adding the normalized galaxy count map scaled by parameter $\lambda$ to the SGWB noise map realizations, as defined by Eq.~\ref{eq:lambda}. The corresponding mean $\Gamma$ and its 5th percentile (obtained over 1000 noise realizations at each $\lambda$ value) are plotted as a function of $\lambda$. The red dashed line denotes the $\Gamma$ value obtained using the O2 data for the same GW band and galaxy catalog. The intersection point between the red dashed line and the 5th percentile curve gives the $\lambda_{95}$ value.
}
\label{fig:signalphoto}
\end{figure}

Figure \ref{fig:signalphoto} shows an example of $\Gamma$ as a function of $\lambda$, computed using the SGWB map in the 50-100 Hz frequency band and the full photometric SDSS catalog. We define $\lambda_{95}$ to be the 95\% confidence upper limit on the scaling factor, i.e. the value of $\lambda$ that yields coherence $\Gamma$ larger than the observed coherence in 95\% of the simulations. For the example shown in Figure~\ref{fig:signalphoto}, we find that $\lambda_{95}=2.7\times 10^{-49}$ st$^{-1}$.
The calculation is repeated for all frequency bands for both photometric and spectroscopic catalogs, and the corresponding $\lambda_{95}$ values are summarized in Table \ref{table:lambda}.
\begin{table}
\begin{tabular}{c |c c c c}
\hline
GW frequency (Hz)&50-100&100-150&150-200&200-250\\
\hline
GC photoz&2.7e-49&1.0e-50&1.7e-47&2.1e-47\\
GC specz&1.1e-49&1.0e-48&1.0e-47&1.2e-46\\
\hline
\end{tabular}
\caption{$\lambda_{95}$ values in units of st$^{-1}$, for the four GW frequency bands and for the photometric and spectroscopic catalogs assuming the full redshift ranges.}
\label{table:lambda}
\end{table}
\section{Discussion and Conclusions}\label{sec:conclusion}
Studying the cross correlations between the SGWB and EM tracers of matter structure offers both the possibility of detecting the SGWB anisotropy sooner and the possibility to probe cosmological and astrophysical parameters driving the formation of structure. In this paper we have laid out a formalism to measure such SGWB-EM correlations. We have used the LIGO data from the second observing run and the galaxy catalog data from the SDSS to study the correlations of different GW frequency bands and different redshift slices in galaxy catalogs. We found no evidence for correlations between the SGWB and galaxy catalogs in these data. 

We emphasize that while this may be the first measurement of its kind, there are many possible directions that should be explored in future works. We outline some of the possibilities here:
\begin{itemize}
\item Our work has used only galaxy counts to track the matter structure. This can be expanded to use weak lensing survey data, or the cosmic microwave and infrared background data (e.g. from Planck~\citep{Aghanim:2019ame,Ade:2013zsi}), or the X-ray data measured by Chandra X-ray Observatory \citep{Schwartz:2004zb}. Different EM tracers will potentially correlate with different components of the SGWB: for example, galaxy counts or weak lensing may correlate with BBH/BNS SGWB, while the CMB anisotropy may correlate with cosmological SGWB models. Hence, spatial correlations with different EM tracers may help distinguish different SGWB contributions. 
\item Vast amounts of new data are expected in the coming decade, on both GW and EM fronts. Advanced LIGO and Advanced Virgo are soon to complete the third observation run, with a sequence of detector upgrades and additional observation runs being planned. The Euclid~\citep{Paykari:2019emf} and SPHEREx~\citep{SPHEREx} missions are expected to produce unprecedented galaxy surveys. For example, Euclid will identify 3$\times 10^7$ emission line galaxies, and use them as tracers of the large scale structure at $1<z<2.5$, which will significantly expand upon the existing SDSS catalogs. 
\item The recently proposed Bayesian approach to measuring the BBH SGWB~\citep{Smith:2017vfk,Ashton:2019wvo} promises to be significantly more sensitive to this type of background than the traditional stochastic search techniques (used also in this paper). This approach can produce the Bayesian posterior distribution of the BBH sky positions (for the entire BBH population), which can then be used to study correlations with the EM tracers such as galaxy counts or weak lensing surveys. The Bayesian approach also offers the possibility of extracting the redshift distribution of the BBH population, giving rise to the possibility of studying 3D correlations (sky position plus redshift) between the BBH SGWB and the galaxy count catalogs.~\citep{Sharan}
\item Our analysis included estimation of a scaling parameter in a simple empirical model of the correlation between SGWB and galaxy counts. This can be expanded to include more sophisticated models of the BBH/BNS SGWB that properly take into account the cosmological and astrophysical evolution \citep{Cusin:2017fwz, Cusin:2017mjm, Cusin:2018rsq, Cusin:2019jpv, Pitrou:2019rjz}.
\end{itemize}
We conclude by noting that studying the SGWB-EM correlations is a good example of how multi-messenger data can be used to generate new probes of astrophysics and cosmology. Upcoming data sets from both GW and EM detectors and telescopes, combined with improvements in data analysis techniques, promise novel ways of probing the evolution of structure in the universe, and perhaps also models of the early universe. 

\section*{Acknowledgements}
The authors thank Giulia Cusin for numerous discussions and insights regarding this manuscript. This work was supported by the NSF grant PHY-1806630.
The authors are thankful for the computing resources provided by LIGO Laboratory and supported by the National Science Foundation grants PHY–0757058 and PHY–0823459.
The code for the analysis in this paper is available upon request.
This paper is assigned the LIGO document control number LIGO-P2000220.

\section*{Data Availability}
The data that support the findings of this study for the LIGO side are openly available in "O2 Data Release" at \url{https://www.gw-openscience.org/data/}. For the SDSS data, the photometric catalog ("photoObj") is available at \url{https://www.sdss.org/dr16/imaging/catalogs/}, the spectroscopic catalogs are available at \url{https://www.sdss.org/dr16/spectro/}.

\bibliography{GW_GC_corr}
\bibliographystyle{mnras}
\bsp
\end{document}